\documentclass{caosp310}

\usepackage{graphicx}

\usepackage{natbib}
\usepackage{longtable}
\usepackage{booktabs}
\usepackage{url}

\articleNo{}
\pubyear{2026}
\volume{}
\volnumber{}
\firstpage{1}
\received{June 23, 2026}
\accepted{July 6, 2026}

\def\BibTeX{{\rm B\kern-.05em{\sc i\kern-.025em b}\kern-.08em
             T\kern-.1667em\lower.7ex\hbox{E}\kern-.125emX}}

\begin{document}

%
\hauthor{M.\,Kamenec, P.\,Gajdo\v{s}, \v{S}.\,Parimucha}

\title{DSLR times of minima of selected eclipsing binaries}


%
%
\author{
        M.\,Kamenec\orcid{0009-0002-2976-0719}
        \and P.\,Gajdo\v{s}\orcid{0000-0003-1478-3256}
        \and \v{S}.\,Parimucha\orcid{0000-0002-7204-9220}
       }

%
\institute{
           \kosice, \email{matus.kamenec@student.upjs.sk}
          }

\date{March 8, 2003}

\maketitle

\begin{abstract}
We present a collection of eclipse timings of eclipsing binary stars obtained over the extended time interval 2014–2024. In all observations, a DSLR camera was used as the detector. The presented dataset contains 250 previously unpublished minima of 86 eclipsing binary systems obtained at three observing sites. The minima were determined from calibrated light curves using the phenomenological modelling approach, and their uncertainties were estimated by bootstrap resampling. The presented measurements extend the available timing coverage of the observed systems and may contribute to future studies of orbital period variations and other long-term phenomena in eclipsing binaries.
\keywords{eclipsing binaries -- times of minima -- digital single lens reflex -- photometry}
\end{abstract}

%
\section{Introduction}
\label{intr}
The rapid development of digital single-lens reflex (DSLR) cameras since the early 2000s has shown that these instruments are useful not only for astrophotography, but also for photometric measurements of variable stars and exoplanetary transits \citep[e.g.][]{2010JBAA..120..157L, 2016PASP..128c5001Z, 2023OEJV..243....1K}. Several studies have confirmed that DSLR sensors can serve as practical alternatives to dedicated CCD and, more recently, CMOS cameras \citep[e.g.][]{2016PASP..128c5001Z, 2023OEJV..243....1K}.

Times of minima are among the most important observational quantities in the study of eclipsing binaries. Long-term collections of eclipse timings allow us to investigate orbital period changes caused by light-time effects due to additional companions \citep{1952ApJ...116..211I}, apsidal motion in eccentric binaries \citep{1983Ap&SS..92..203G}, dynamical perturbations in multiple systems \citep{2015MNRAS.448..946B}, mass-transfer processes, and magnetic activity cycles \citep{1992ApJ...385..621A}.

In this paper we present 250 DSLR-based times of minima for 86 eclipsing binary systems, obtained between 2014 and 2024. All reported timings are new and have not been published previously. The observations were carried out mainly at the Kolonica Saddle Observatory and at a private observatory in Prešov, with one additional dataset from the observatory in Valašské Meziříčí.
The target list was updated continuously based on target visibility, recommendations from the local variable-star community, and the scientific interest of individual systems. We paid particular attention to eclipsing binaries that are regularly monitored because of suspected or known orbital-period variations reflected in their O–C diagrams. Most of the observed targets are short-period systems, which often made it possible to record both primary and secondary minima in a single night, or to observe minima of several different systems consecutively.
Since our main goal was to determine eclipse timings rather than to model the light curves in detail, DSLR cameras turned out to be well suited for this kind of work. Timing measurements do not require high-precision multi-band photometry — the time of minimum can be derived reliably from the shape of the eclipse profile even with moderate photometric accuracy.

\section{Observation and data reduction}
\label{oadr}

The observations of minima were either planned in advance or obtained as by-products of observations aimed at constructing phase curves and further analysis of selected systems. Observation planning relied on auxiliary tools, namely PozorWin (Pribulla, private communication) and ephemeris predictions available through the International Variable Star Index \citep{VSX}. The instrumental setup evolved over the course of the programme; in total, we used 
two Newtonian reflectors, and one refractor, as summarised in Tab.~\ref{tab:systems}.

\begin{table}[t]
\caption{Telescope systems used for the observations.}
\centering
\begin{tabular}{clc}
\hline\hline
System & Telescope & Mount \\
\hline
I & Newton 150/750\,mm & EQ3-2 GoTo \\
II & Apochromatic refractor 102/500\,mm & EQ3-2 GoTo \\
III & Newton 200/1200\,mm & NEQ-6 \\
\hline\hline
\end{tabular}

\label{tab:systems}

\end{table}
The observing sites are listed in Tab.~\ref{tab:sites}. Kolonica Saddle lies within the Poloniny Dark Sky Park and offers genuinely dark skies, while Pre\v{s}ov and  Vala\v{s}sk\'{e} Mezi\v{r}\'{i}\v{c}\'{i} are situated in urban residential areas and are affected by moderate light pollution.

\begin{table}[t]
\caption{Observing sites.}
{\centering
\footnotesize
\begin{tabular}{cccccccc}
\hline\hline
Site ID & Site & Country & Coordinates & Alt. (m) & SQM & Bortle class\\
\hline
KOL & Kolonica Saddle & Slovakia & 48.936N, 22.274E & 441 & 21.89 & 3 \\
PO & Pre\v{s}ov & Slovakia & 48.974N, 21.261E & 248 & 20.26 & 5 \\
VM & Vala\v{s}sk\'{e} Mezi\v{r}\'{i}\v{c}\'{i} & Czechia & 49.463N, 17.973E & 335 & 20.73 & 4 \\
\hline\hline
\end{tabular}}
\\
\textit{Note.}
Sky Quality Meter (SQM) values in $\mathrm{mag\,arcsec^{-2}}$ and Bortle class values taken from \url{https://www.lightpollutionmap.info/}
\label{tab:sites}

\end{table}

All observations were made with an uncooled Canon EOS 500D DSLR camera, equipped with a 15.1-megapixel APS-C CMOS sensor measuring 22.3 $\times$ 14.9~mm. The detector has a pixel array of 4752 $\times$ 3168 pixels with a pixel size of 4.69~$\mu$m and 14-bit analogue-to-digital conversion. Like most DSLR sensors, it is covered by a Bayer filter array of red, green, and blue filters placed over individual pixels. Since this filter array cannot be removed, all photometry was performed in the so-called grayscale (or ``clear'') mode, in which pixel values are recorded regardless of their Bayer-filter colour, allowing the full sensor area to contribute to the photometric signal \citep{2023OEJV..243....1K}.

Both mount control and image acquisition were managed from a computer. ISO settings and exposure times were adjusted for each target to avoid saturation. The observed stars typically fell in the brightness range 10--12~mag, for which ISO settings of 800--1600 and an exposure time of 30~seconds were used. Images were acquired using one of three software packages --- MaximDL\footnote{\url{https://diffractionlimited.com/product/maxim-dl/}}, Astrophotography Tool\footnote{\url{https://astrophotography.app/}}, or CCDCiel\footnote{\url{https://www.ap-i.net/ccdciel/en/start}} --- and stored either in native RAW format or directly as FITS files. The computer clock was synchronized using the \mbox{NetTime}\footnote{\url{https://www.timesynctool.com/}} software. The acquisition software recorded the UTC start time of each exposure with sub-second precision.

Image processing and calibration were carried out with the Muniwin software package\footnote{\url{https://c-munipack.sourceforge.net/}} \citep{2014ascl.soft02006H}, which natively supports both file formats. All images were calibrated using dark frames and flat fields obtained with the same ISO settings and exposure times as the corresponding science frames; dark frames were recorded at temperatures closely matching those during the observations. As a final step of the reduction, timestamps were converted to Heliocentric Julian Dates (HJD). Standard aperture photometry was then performed, comparison and check stars were selected, and the resulting light curve was constructed and exported for further analysis.

\section{Minima times determination}
\label{tomd}

The times of minima were determined from the observed light curves using the phenomenological modelling approach of \citet{2015A&A...584A...8M}. In this method, the eclipse profile is approximated by an analytical function that captures the observed shape of the minimum, and the time of minimum is derived as the model parameter corresponding to the centre of symmetry of the eclipse. Because the fit uses the full eclipse profile rather than only observations near the bottom of the minimum, the resulting timings are generally less affected by observational scatter and incomplete phase coverage. The phenomenological model describes the eclipse shape reliably without requiring any physical modelling of the binary system, and the minimum time follows directly from the fitted profile. This makes the approach both practical and well-suited to datasets of varying photometric quality.

Uncertainties were estimated using a bootstrap resampling procedure following \citet{2012A&A...544L...3C}. In contrast to analytical error estimates, the bootstrap method derives timing uncertainty directly from the data and therefore naturally reflects the actual noise level, sampling cadence, and photometric quality of each light curve. This generally leads to more realistic uncertainty estimates for eclipse timings \citep{2012A&A...544L...3C, 10.1214/aos/1176344552, bootstrap}. All minima presented in Tab.~\ref{tab:minima} were determined using the same fitting and uncertainty estimation procedure, which ensures consistency across the entire dataset.

\begin{longtable}{lccccc}
\caption{DSLR times of minima.}
\label{tab:minima} \\

\toprule
HJD & $\sigma_+$ & $\sigma_-$ & Type & Site ID & System \\
\midrule
\endfirsthead
\toprule
HJD & $\sigma_+$ & $\sigma_-$ & Type & Site ID & System \\
\midrule
\endhead
\multicolumn{6}{l}{\textbf{AB And}}\\[0.3ex]
2456858.44583 & 0.00014 & 0.00014 & p & KOL & I \\
2457577.48605 & 0.00015 & 0.00015 & s & KOL & I \\
2458133.23380 & 0.00010 & 0.00009 & p & PO & II \\
\multicolumn{6}{l}{\textbf{AD And}}\\[0.3ex]
2459873.23762 & 0.00028 & 0.00027 & s & KOL & III \\
\multicolumn{6}{l}{\textbf{BD And}}\\[0.3ex]
2459811.47406 & 0.00025 & 0.00024 & s & KOL & III \\
\multicolumn{6}{l}{\textbf{EP And}}\\[0.3ex]
2458354.35106 & 0.00050 & 0.00127 & p & KOL & I \\
2459880.26265 & 0.00023 & 0.00021 & p & KOL & III \\
\multicolumn{6}{l}{\textbf{GZ And}}\\[0.3ex]
2457977.45556 & 0.00026 & 0.00028 & s & KOL & I \\
2458532.27716 & 0.00031 & 0.00022 & s & PO & II \\
\multicolumn{6}{l}{\textbf{LO And}}\\[0.3ex]
2458353.33715 & 0.00051 & 0.00018 & p & KOL & I \\
\multicolumn{6}{l}{\textbf{RT And}}\\[0.3ex]
2456864.48750 & 0.00011 & 0.00012 & p & KOL & I \\
2458133.35155 & 0.00051 & 0.00050 & s & PO & II \\
2459749.38491 & 0.00013 & 0.00012 & p & KOL & III \\
\multicolumn{6}{l}{\textbf{V483 And}}\\[0.3ex]
2459785.44377 & 0.00022 & 0.00025 & p & KOL & III \\
2459818.41690 & 0.00028 & 0.00029 & p & KOL & III \\
\multicolumn{6}{l}{\textbf{AC Boo}}\\[0.3ex]
2458379.35665 & 0.00080 & 0.00043 & p & PO & II \\
2459782.46884 & 0.00028 & 0.00024 & p & PO & II \\
2459783.52443 & 0.00057 & 0.00056 & p & PO & II \\
2460070.42200 & 0.00008 & 0.00010 & p & KOL & III \\
2460171.38856 & 0.00043 & 0.00076 & s & PO & II \\
\multicolumn{6}{l}{\textbf{TY Boo}}\\[0.3ex]
2459420.34065 & 0.00129 & 0.00099 & p & PO & II \\
2460192.27537 & 0.00021 & 0.00018 & p & PO & II \\
\multicolumn{6}{l}{\textbf{TZ Boo}}\\[0.3ex]
2457994.39412 & 0.00024 & 0.00024 & s & KOL & I \\
2458360.35199 & 0.00038 & 0.00116 & p & PO & II \\
2459742.44251 & 0.00021 & 0.00020 & p & KOL & III \\
2460170.35273 & 0.00038 & 0.00035 & p & PO & II \\
2460175.40441 & 0.00036 & 0.00035 & p & PO & II \\
2460181.34458 & 0.00044 & 0.00046 & p & PO & II \\
\multicolumn{6}{l}{\textbf{V339 Boo}}\\[0.3ex]
2459748.41925 & 0.00040 & 0.00041 & p & KOL & III \\
2459750.41184 & 0.00054 & 0.00046 & s & KOL & III \\
\multicolumn{6}{l}{\textbf{AO Cam}}\\[0.3ex]
2458218.28395 & 0.00023 & 0.00024 & p & PO & II \\
2458230.32697 & 0.00041 & 0.00059 & s & PO & II \\
2458532.34742 & 0.00005 & 0.00079 & p & PO & II \\
2459656.31230 & 0.00016 & 0.00015 & p & PO & II \\
2459657.30118 & 0.00017 & 0.00017 & p & PO & II \\
2460005.34342 & 0.00013 & 0.00013 & p & PO & II \\
\multicolumn{6}{l}{\textbf{AZ Cam}}\\[0.3ex]
2459503.28942 & 0.00047 & 0.00045 & p & PO & II \\
\multicolumn{6}{l}{\textbf{CD Cam}}\\[0.3ex]
2456746.39357 & 0.00056 & 0.00056 & p & KOL & I \\
\multicolumn{6}{l}{\textbf{DN Cam}}\\[0.3ex]
2459664.38402 & 0.00018 & 0.00017 & s & PO & II \\
\multicolumn{6}{l}{\textbf{HW Cam}}\\[0.3ex]
2459731.45795 & 0.00029 & 0.00026 & p & PO & II \\
2459743.46873 & 0.00046 & 0.00045 & p & PO & II \\
\multicolumn{6}{l}{\textbf{NR Cam}}\\[0.3ex]
2458223.45928 & 0.00030 & 0.00031 & s & PO & II \\
2458223.32773 & 0.00021 & 0.00022 & p & PO & II \\
2458348.45628 & 0.00004 & 0.00052 & p & PO & I \\
2459652.45368 & 0.00025 & 0.00029 & p & PO & II \\
2459652.57780 & 0.00034 & 0.00038 & s & PO & II \\
2459656.41577 & 0.00031 & 0.00030 & s & PO & II \\
2459657.43953 & 0.00038 & 0.00039 & s & PO & II \\
2460004.55036 & 0.00018 & 0.00021 & p & KOL & III \\
2460085.53583 & 0.00024 & 0.00026 & s & KOL & III \\
2460085.40876 & 0.00064 & 0.00086 & p & KOL & III \\
2460408.46246 & 0.00022 & 0.00021 & s & KOL & III \\
2460408.59109 & 0.00021 & 0.00019 & p & KOL & III \\
\multicolumn{6}{l}{\textbf{NU Cam}}\\[0.3ex]
2459659.53924 & 0.00090 & 0.00094 & s & PO & II \\
2459660.57469 & 0.00055 & 0.00052 & p & PO & II \\
\multicolumn{6}{l}{\textbf{SV Cam}}\\[0.3ex]
2457240.53436 & 0.00020 & 0.00019 & p & KOL & I \\
2457963.48495 & 0.00036 & 0.00035 & p & KOL & I \\
2459595.62033 & 0.00010 & 0.00010 & p & PO & II \\
\multicolumn{6}{l}{\textbf{V584 Cam}}\\[0.3ex]
2459595.41453 & 0.00053 & 0.00050 & p & PO & II \\
\multicolumn{6}{l}{\textbf{V608 Cam}}\\[0.3ex]
2459681.30569 & 0.00070 & 0.00072 & s & PO & II \\
2459684.44106 & 0.00074 & 0.00071 & s & PO & II \\
2459736.41993 & 0.00101 & 0.00136 & s & PO & II \\
2459758.37596 & 0.00042 & 0.00047 & s & KOL & III \\
2459796.45937 & 0.00042 & 0.00042 & s & KOL & III \\
2459871.28252 & 0.00058 & 0.00060 & s & KOL & III \\
2459912.28431 & 0.00010 & 0.00010 & p & KOL & III \\
2460086.35476 & 0.00031 & 0.00035 & s & KOL & III \\
\multicolumn{6}{l}{\textbf{BS Cas}}\\[0.3ex]
2459797.47612 & 0.00015 & 0.00017 & p & KOL & III \\
2460134.42994 & 0.00014 & 0.00013 & p & KOL & III \\
2460171.42813 & 0.00020 & 0.00018 & p & KOL & III \\
\multicolumn{6}{l}{\textbf{CW Cas}}\\[0.3ex]
2458352.34338 & 0.00042 & 0.00004 & p & KOL & I \\
2459595.24635 & 0.00016 & 0.00016 & p & PO & II \\
2459817.33219 & 0.00011 & 0.00013 & s & KOL & III \\
2459817.49015 & 0.00008 & 0.00008 & p & KOL & III \\
2459968.30968 & 0.00013 & 0.00013 & p & KOL & III \\
2460135.39130 & 0.00024 & 0.00022 & p & KOL & III \\
\multicolumn{6}{l}{\textbf{V523 Cas}}\\[0.3ex]
2457581.45014 & 0.00010 & 0.00011 & s & KOL & I \\
2457965.52716 & 0.00007 & 0.00007 & p & KOL & I \\
2458531.30138 & 0.00035 & 0.00016 & p & PO & II \\
2459510.47663 & 0.00016 & 0.00016 & p & KOL & III \\
2459816.49977 & 0.00008 & 0.00008 & s & KOL & III \\
2459816.38322 & 0.00007 & 0.00008 & p & KOL & III \\
2459821.52463 & 0.00010 & 0.00011 & p & KOL & III \\
2459821.40714 & 0.00008 & 0.00007 & s & KOL & III \\
2459828.30184 & 0.00007 & 0.00007 & p & KOL & III \\
2459828.41811 & 0.00007 & 0.00007 & s & KOL & III \\
2460141.45233 & 0.00006 & 0.00006 & p & KOL & III \\
2460193.33312 & 0.00007 & 0.00008 & p & KOL & III \\
\multicolumn{6}{l}{\textbf{V651 Cas}}\\[0.3ex]
2459513.44400 & 0.00007 & 0.00009 & p & PO & II \\
\multicolumn{6}{l}{\textbf{V1297 Cas}}\\[0.3ex]
2459783.44089 & 0.00020 & 0.00022 & p & KOL & III \\
\multicolumn{6}{l}{\textbf{EF Cep}}\\[0.3ex]
2459780.42807 & 0.00019 & 0.00019 & p & KOL & III \\
\multicolumn{6}{l}{\textbf{GW Cep}}\\[0.3ex]
2457964.41394 & 0.00019 & 0.00020 & p & KOL & I \\
2459652.31220 & 0.00024 & 0.00025 & p & PO & II \\
2459653.42913 & 0.00038 & 0.00040 & s & PO & II \\
2459746.36691 & 0.00012 & 0.00011 & p & KOL & III \\
2459797.38000 & 0.00009 & 0.00008 & p & KOL & III \\
2459990.27421 & 0.00017 & 0.00025 & p & PO & II \\
2459991.23055 & 0.00020 & 0.00026 & p & PO & II \\
2460149.37097 & 0.00017 & 0.00016 & p & KOL & III \\
\multicolumn{6}{l}{\textbf{OT Cep}}\\[0.3ex]
2459745.43858 & 0.00014 & 0.00015 & p & KOL & III \\
\multicolumn{6}{l}{\textbf{VW Cep}}\\[0.3ex]
2459519.40644 & 0.00028 & 0.00030 & s & PO & II \\
2459519.26540 & 0.00000 & 0.00056 & p & PO & II \\
\multicolumn{6}{l}{\textbf{WZ Cep}}\\[0.3ex]
2457973.42104 & 0.00021 & 0.00022 & s & KOL & I \\
\multicolumn{6}{l}{\textbf{V902 Cep}}\\[0.3ex]
2459516.52088 & 0.00040 & 0.00039 & p & PO & II \\
2459517.50695 & 0.00033 & 0.00032 & p & PO & II \\
2459534.27012 & 0.00068 & 0.00073 & p & PO & II \\
2459535.26110 & 0.00054 & 0.00051 & p & PO & II \\
2459562.37442 & 0.00059 & 0.00053 & s & PO & II \\
2460065.45629 & 0.00028 & 0.00024 & p & KOL & III \\
2460143.52164 & 0.00109 & 0.00106 & s & KOL & III \\
\multicolumn{6}{l}{\textbf{V955 Cep}}\\[0.3ex]
2459502.56360 & 0.00072 & 0.00073 & p & PO & II \\
\multicolumn{6}{l}{\textbf{TX Cnc}}\\[0.3ex]
2458227.39257 & 0.00047 & 0.00046 & s & PO & II \\
2460410.39161 & 0.00031 & 0.00032 & p & KOL & III \\
\multicolumn{6}{l}{\textbf{CC Com}}\\[0.3ex]
2459658.58499 & 0.00026 & 0.00025 & p & PO & II \\
2460006.38161 & 0.00011 & 0.00011 & p & KOL & III \\
2460020.28473 & 0.00010 & 0.00012 & p & KOL & III \\
\multicolumn{6}{l}{\textbf{RW Com}}\\[0.3ex]
2459750.39162 & 0.00022 & 0.00022 & p & PO & II \\
2460100.47968 & 0.00020 & 0.00019 & p & KOL & III \\
\multicolumn{6}{l}{\textbf{RZ Com}}\\[0.3ex]
2459403.37043 & 0.00024 & 0.00024 & s & PO & II \\
\multicolumn{6}{l}{\textbf{YY CrB}}\\[0.3ex]
2457996.38787 & 0.00028 & 0.00025 & p & KOL & I \\
\multicolumn{6}{l}{\textbf{BI CVn}}\\[0.3ex]
2459402.41870 & 0.00057 & 0.00136 & p & PO & II \\
2460122.43049 & 0.00025 & 0.00025 & p & KOL & III \\
\multicolumn{6}{l}{\textbf{DL CVn}}\\[0.3ex]
2459728.43894 & 0.00079 & 0.00072 & p & PO & II \\
\multicolumn{6}{l}{\textbf{CG Cyg}}\\[0.3ex]
2459516.31537 & 0.00011 & 0.00012 & p & PO & II \\
2460176.49246 & 0.00008 & 0.00009 & p & KOL & III \\
2460177.44011 & 0.00014 & 0.00013 & s & KOL & III \\
2460178.38626 & 0.00011 & 0.00012 & p & KOL & III \\
2460181.54189 & 0.00006 & 0.00007 & p & KOL & III \\
\multicolumn{6}{l}{\textbf{V401 Cyg}}\\[0.3ex]
2460168.48169 & 0.00017 & 0.00014 & p & KOL & III \\
\multicolumn{6}{l}{\textbf{V1018 Cyg}}\\[0.3ex]
2459486.45833 & 0.00096 & 0.00096 & p & PO & II \\
\multicolumn{6}{l}{\textbf{V1191 Cyg}}\\[0.3ex]
2457964.52597 & 0.00023 & 0.00025 & p & KOL & I \\
\multicolumn{6}{l}{\textbf{V1918 Cyg}}\\[0.3ex]
2456865.47915 & 0.00017 & 0.00021 & p & KOL & I \\
2457957.50619 & 0.00021 & 0.00021 & p & KOL & I \\
2459512.29336 & 0.00027 & 0.00025 & p & PO & II \\
2459517.24954 & 0.00020 & 0.00018 & p & PO & II \\
2459866.38306 & 0.00022 & 0.00019 & p & PO & II \\
2460075.45112 & 0.00010 & 0.00011 & p & KOL & III \\
\multicolumn{6}{l}{\textbf{V2277 Cyg}}\\[0.3ex]
2459502.37053 & 0.00015 & 0.00020 & p & PO & II \\
\multicolumn{6}{l}{\textbf{AV Del}}\\[0.3ex]
2459754.39392 & 0.00162 & 0.00152 & p & KOL & III \\
2459781.36786 & 0.00100 & 0.00109 & p & KOL & III \\
\multicolumn{6}{l}{\textbf{AX Dra}}\\[0.3ex]
2460428.50409 & 0.00009 & 0.00009 & p & KOL & III \\
2460429.35302 & 0.00053 & 0.00117 & s & KOL & III \\
2460435.32036 & 0.00011 & 0.00012 & p & KOL & III \\
\multicolumn{6}{l}{\textbf{BE Dra}}\\[0.3ex]
2457967.48118 & 0.00032 & 0.00033 & p & KOL & I \\
2458382.34409 & 0.00008 & 0.00086 & p & PO & II \\
2460445.42959 & 0.00038 & 0.00036 & s & KOL & III \\
2460446.47254 & 0.00028 & 0.00028 & s & KOL & III \\
\multicolumn{6}{l}{\textbf{EF Dra}}\\[0.3ex]
2457975.45834 & 0.00040 & 0.00038 & p & KOL & I \\
2459826.35536 & 0.00039 & 0.00036 & p & PO & II \\
2460020.56310 & 0.00041 & 0.00045 & p & KOL & III \\
2460181.48457 & 0.00037 & 0.00040 & s & KOL & III \\
2460184.45217 & 0.00069 & 0.00073 & s & KOL & III \\
2460199.50290 & 0.00041 & 0.00046 & p & PO & II \\
\multicolumn{6}{l}{\textbf{FU Dra}}\\[0.3ex]
2457582.46906 & 0.00011 & 0.00011 & p & KOL & I \\
2457987.34091 & 0.00015 & 0.00015 & p & KOL & I \\
2457990.40826 & 0.00014 & 0.00014 & p & KOL & I \\
2458371.35560 & 0.00056 & 0.00047 & p & PO & II \\
2458378.40967 & 0.00062 & 0.00016 & p & PO & II \\
2459402.54875 & 0.00111 & 0.00105 & p & PO & II \\
2459403.46734 & 0.00116 & 0.00050 & p & PO & II \\
2459404.54407 & 0.00046 & 0.00053 & s & PO & II \\
2459404.38927 & 0.00051 & 0.00051 & p & PO & II \\
2459824.44214 & 0.00017 & 0.00025 & s & PO & II \\
2460007.39824 & 0.00018 & 0.00018 & p & KOL & III \\
2460137.44779 & 0.00008 & 0.00008 & p & KOL & III \\
2460184.37482 & 0.00041 & 0.00039 & p & KOL & III \\
\multicolumn{6}{l}{\textbf{HL Dra}}\\[0.3ex]
2459513.27802 & 0.00055 & 0.00063 & p & PO & II \\
\multicolumn{6}{l}{\textbf{MZ Dra}}\\[0.3ex]
2460320.30591 & 0.00023 & 0.00022 & p & KOL & III \\
\multicolumn{6}{l}{\textbf{V380 Dra}}\\[0.3ex]
2459452.48806 & 0.00039 & 0.00049 & s & PO & II \\
2459461.37468 & 0.00031 & 0.00029 & s & PO & II \\
2459462.36228 & 0.00059 & 0.00057 & s & PO & II \\
2459466.31405 & 0.00029 & 0.00027 & s & PO & II \\
2459484.33453 & 0.00020 & 0.00019 & p & PO & II \\
2459835.38068 & 0.00024 & 0.00022 & p & PO & II \\
2459869.44827 & 0.00009 & 0.00016 & p & KOL & I \\
\multicolumn{6}{l}{\textbf{V405 Dra}}\\[0.3ex]
2460443.45278 & 0.00018 & 0.00018 & s & KOL & III \\
\multicolumn{6}{l}{\textbf{V728 Her}}\\[0.3ex]
2459493.29475 & 0.00054 & 0.00056 & p & PO & II \\
2459502.25215 & 0.00051 & 0.00050 & p & PO & II \\
2459802.47311 & 0.00062 & 0.00066 & p & PO & II \\
2460171.49534 & 0.00056 & 0.00052 & p & PO & II \\
\multicolumn{6}{l}{\textbf{V829 Her}}\\[0.3ex]
2457984.42564 & 0.00035 & 0.00029 & s & KOL & I \\
2459799.39132 & 0.00069 & 0.00068 & p & KOL & III \\
2460169.36116 & 0.00050 & 0.00039 & p & PO & II \\
2460170.43614 & 0.00037 & 0.00035 & p & PO & II \\
2460193.35769 & 0.00044 & 0.00044 & p & PO & II \\
\multicolumn{6}{l}{\textbf{V857 Her}}\\[0.3ex]
2459504.27549 & 0.00064 & 0.00047 & p & PO & II \\
2459779.48032 & 0.00033 & 0.00035 & p & PO & II \\
\multicolumn{6}{l}{\textbf{PP Lac}}\\[0.3ex]
2458350.35329 & 0.00025 & 0.00020 & p & KOL & I \\
2459504.50295 & 0.00022 & 0.00021 & p & PO & II \\
2460097.42480 & 0.00015 & 0.00016 & p & KOL & III \\
\multicolumn{6}{l}{\textbf{SW Lac}}\\[0.3ex]
2458134.25625 & 0.00011 & 0.00012 & p & PO & II \\
2459512.38900 & 0.00006 & 0.00005 & p & PO & II \\
\multicolumn{6}{l}{\textbf{V344 Lac}}\\[0.3ex]
2459517.35800 & 0.00047 & 0.00050 & p & PO & II \\
\multicolumn{6}{l}{\textbf{CE Leo}}\\[0.3ex]
2459726.39758 & 0.00033 & 0.00034 & p & PO & II \\
\multicolumn{6}{l}{\textbf{UV Leo}}\\[0.3ex]
2458251.37130 & 0.00025 & 0.00029 & p & PO & II \\
\multicolumn{6}{l}{\textbf{TZ Lyr}}\\[0.3ex]
2459811.55309 & 0.00074 & 0.00088 & p & PO & II \\
2459870.25284 & 0.00017 & 0.00018 & p & PO & II \\
2459871.31030 & 0.00015 & 0.00014 & p & PO & II \\
\multicolumn{6}{l}{\textbf{UV Lyn}}\\[0.3ex]
2459667.41873 & 0.00038 & 0.00037 & p & PO & II \\
\multicolumn{6}{l}{\textbf{V508 Oph}}\\[0.3ex]
2458377.35657 & 0.00031 & 0.00053 & p & PO & II \\
2459441.38432 & 0.00006 & 0.00006 & p & VM & III \\
\multicolumn{6}{l}{\textbf{BX Peg}}\\[0.3ex]
2456865.37891 & 0.00013 & 0.00013 & p & KOL & I \\
2457234.40818 & 0.00030 & 0.00028 & p & KOL & I \\
2457580.44201 & 0.00016 & 0.00016 & p & KOL & I \\
2457939.51836 & 0.00014 & 0.00017 & s & KOL & I \\
2458349.34688 & 0.00017 & 0.00029 & p & KOL & I \\
2459504.38520 & 0.00014 & 0.00015 & p & PO & II \\
2459513.35839 & 0.00013 & 0.00018 & p & PO & II \\
\multicolumn{6}{l}{\textbf{V432 Per}}\\[0.3ex]
2458531.39429 & 0.00054 & 0.00019 & p & PO & II \\
2460194.56876 & 0.00010 & 0.00009 & p & KOL & III \\
\multicolumn{6}{l}{\textbf{CW Sge}}\\[0.3ex]
2460193.48318 & 0.00074 & 0.00073 & p & PO & II \\
\multicolumn{6}{l}{\textbf{AU Ser}}\\[0.3ex]
2456867.37266 & 0.00012 & 0.00013 & p & KOL & I \\
2458001.35281 & 0.00062 & 0.00044 & p & KOL & I \\
2458374.32056 & 0.00030 & 0.00034 & p & PO & II \\
2458375.28801 & 0.00019 & 0.00040 & s & PO & II \\
2459763.38642 & 0.00011 & 0.00010 & p & KOL & III \\
2459767.44550 & 0.00022 & 0.00047 & s & PO & II \\
2460057.31694 & 0.00019 & 0.00015 & s & KOL & III \\
\multicolumn{6}{l}{\textbf{OU Ser}}\\[0.3ex]
2457991.31317 & 0.00046 & 0.00039 & s & KOL & I \\
\multicolumn{6}{l}{\textbf{AH Tau}}\\[0.3ex]
2458566.28967 & 0.00009 & 0.00028 & p & PO & II \\
\multicolumn{6}{l}{\textbf{V781 Tau}}\\[0.3ex]
2459666.26558 & 0.00023 & 0.00022 & p & PO & II \\
2459667.29945 & 0.00024 & 0.00024 & p & PO & II \\
\multicolumn{6}{l}{\textbf{BX Tri}}\\[0.3ex]
2459519.61286 & 0.00127 & 0.00146 & p & PO & II \\
2459824.55583 & 0.00033 & 0.00031 & p & KOL & III \\
2459824.45336 & 0.00066 & 0.00075 & s & KOL & III \\
2459826.48193 & 0.00048 & 0.00038 & p & KOL & III \\
2459826.38027 & 0.00055 & 0.00067 & s & KOL & III \\
2459870.29992 & 0.00065 & 0.00070 & s & KOL & III \\
2459870.40264 & 0.00058 & 0.00048 & p & KOL & III \\
\multicolumn{6}{l}{\textbf{W UMa}}\\[0.3ex]
2459655.60271 & 0.00010 & 0.00009 & p & PO & II \\
\multicolumn{6}{l}{\textbf{AA UMa}}\\[0.3ex]
2458230.43093 & 0.00023 & 0.00026 & p & PO & II \\
2459664.54883 & 0.00028 & 0.00028 & s & PO & II \\
2460006.51446 & 0.00033 & 0.00019 & p & KOL & III \\
\multicolumn{6}{l}{\textbf{ES UMa}}\\[0.3ex]
2460176.40850 & 0.00085 & 0.00129 & p & KOL & III \\
\multicolumn{6}{l}{\textbf{HH UMa}}\\[0.3ex]
2458229.47158 & 0.00072 & 0.00069 & p & PO & II \\
2460066.37353 & 0.00069 & 0.00076 & p & KOL & III \\
\multicolumn{6}{l}{\textbf{UX UMa}}\\[0.3ex]
2459733.37571 & 0.00035 & 0.00026 & p & PO & II \\
2459748.51924 & 0.00017 & 0.00015 & p & PO & II \\
2460099.38049 & 0.00014 & 0.00012 & p & KOL & III \\
\multicolumn{6}{l}{\textbf{VV UMa}}\\[0.3ex]
2460435.46899 & 0.00120 & 0.00150 & s & PO & II \\
2460443.37132 & 0.00024 & 0.00023 & p & PO & II \\
2460445.43294 & 0.00012 & 0.00012 & p & PO & II \\
\multicolumn{6}{l}{\textbf{XY UMa}}\\[0.3ex]
2458218.34547 & 0.00019 & 0.00021 & p & PO & II \\
2458229.36290 & 0.00013 & 0.00014 & p & PO & II \\
2459666.59436 & 0.00061 & 0.00059 & s & PO & II \\
\multicolumn{6}{l}{\textbf{TV UMi}}\\[0.3ex]
2457991.43294 & 0.00174 & 0.00132 & p & KOL & I \\
\multicolumn{6}{l}{\textbf{TY UMi}}\\[0.3ex]
2459803.37993 & 0.00052 & 0.00052 & p & PO & II \\
\multicolumn{6}{l}{\textbf{VW UMi}}\\[0.3ex]
2459782.48803 & 0.00024 & 0.00021 & p & KOL & III \\
\multicolumn{6}{l}{\textbf{NSVS 5789962}}\\[0.3ex]
2459461.46165 & 0.00174 & 0.00399 & p & PO & II \\
2459466.56294 & 0.00169 & 0.00464 & s & PO & II \\
\bottomrule
\end{longtable}

\vspace{0.5ex}
\noindent
\textbf{Notes.}
p and s denote primary and secondary minima, respectively.
The columns Site ID and System correspond to Tab. \ref{tab:sites} and \ref{tab:systems}, respectively.

\section{Limitations}

Several sources of measurement uncertainty are inherent to the use of a DSLR camera as a photometric detector. These include the presence of the Bayer filter array, a lower bit depth compared to dedicated CCD and CMOS cameras, and the absence of sensor cooling, which leads to higher dark current and increased thermal noise — particularly during observations at elevated ambient temperatures. The quantum efficiency of DSLR sensors is also generally lower than that of purpose-built astronomical detectors. Beyond the instrument itself, the observations were acquired under varying conditions, including sites affected by moderate light pollution and changing atmospheric transparency. The observations were also affected by varying atmospheric seeing, although it was not monitored systematically and did not represent a significant limitation for the relatively bright targets included in this study.

\begin{figure}[t]
	\centering
	\includegraphics[width=1.0\columnwidth]{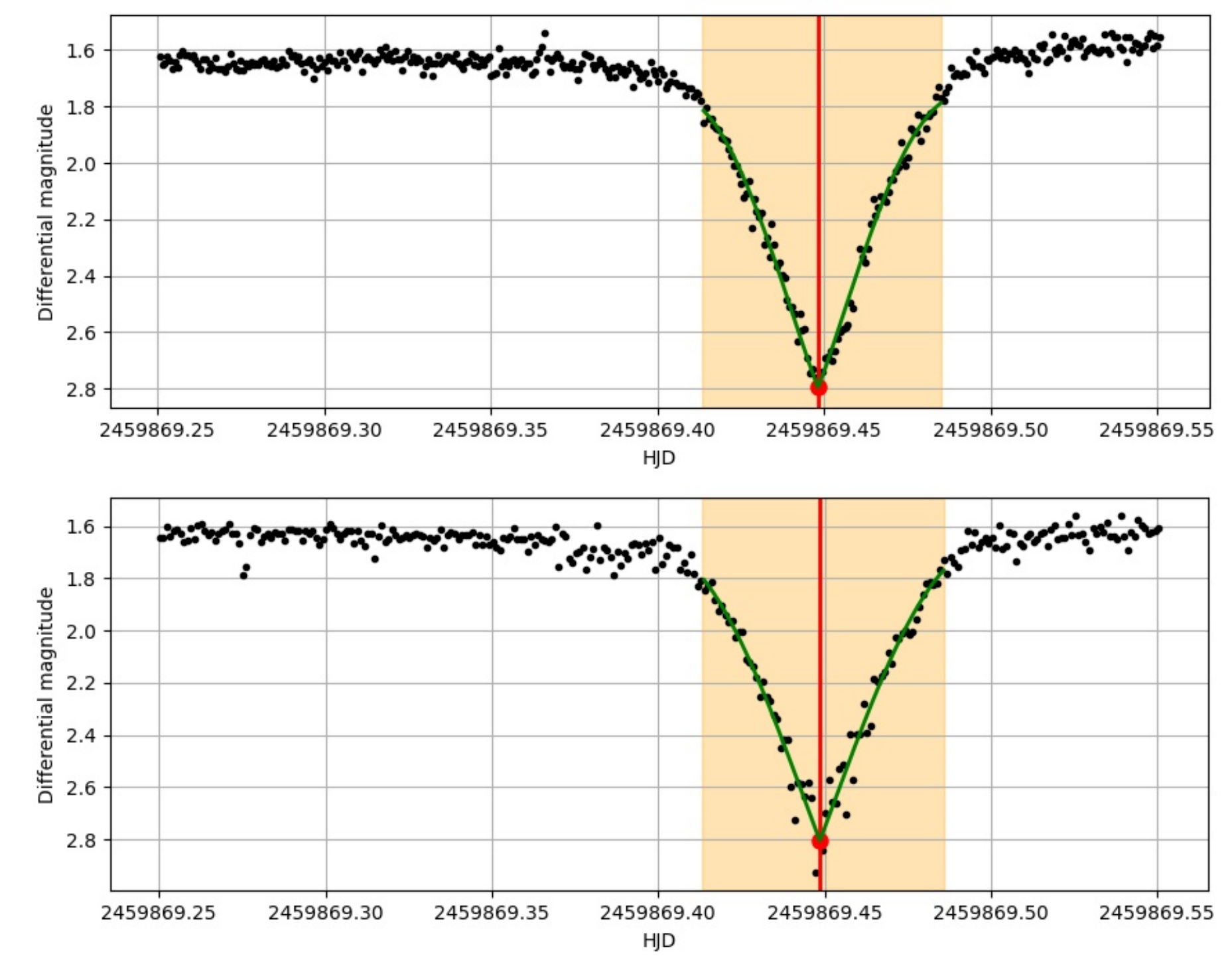}
	\caption{Two lightcurves of V0380 Dra with fitted minima observed simultaneously from Kolonica saddle (above) and from Prešov (bottom).}
	\label{minima}
\end{figure}

To illustrate the effect of observing conditions on DSLR photometry, the same minimum of V380~Dra was observed simultaneously from two sites using different telescope systems: one from Kolonica Saddle within the Dark Sky Park Poloniny, and the other from the urban environment of Prešov (Fig.~\ref{minima}).

As expected, the light curve from the dark-sky site shows noticeably lower point-to-point scatter and a smoother eclipse profile. The urban observation, by contrast, exhibits increased photometric noise, reflecting the combined effect of a brighter sky background, a smaller telescope aperture, and higher detector gain settings. These differences are visible both in the precision of individual measurements and in the uncertainties of the derived eclipse timings (Tab.~\ref{tab:v380dra_comparison}).

Despite the substantially different observing conditions, both datasets produced mutually consistent times of minimum. This comparison shows that DSLR-based eclipse timing remains feasible even in the presence of significant light pollution. While dark-sky observations naturally yield better photometric quality and more precise timing determinations, urban sites can still contribute scientifically useful eclipse timings suitable for long-term period monitoring of eclipsing binaries. For our final compilation, we adopted the timing value derived from the Kolonica Saddle observations.

\begin{table}[t]
\centering
\caption{Comparison of two independent observations of the eclipse minimum of V380 Dra obtained under different observing conditions.}
\label{tab:v380dra_comparison}
\begin{tabular}{lcc}
\hline\hline
Parameter & Kolonica Saddle & Prešov \\
\hline
Site type & Dark-sky site & Urban site \\
SQM ($\mathrm{mag\,arcsec^{-2}}$) & 21.89 & 20.26 \\
Telescope & Newton 200/1200 & APO 102/500 \\
Exposure time (s) & 30 & 60 \\
ISO & 800 & 1800 \\
Point error (mag) & 0.005--0.008 & 0.013--0.016 \\
HJD minimum & 2459869.448266 & 2459869.448703 \\
$\sigma_-$ (d) & 0.0000933 & 0.0002160 \\
$\sigma_+$ (d) & 0.0001567 & 0.0001904 \\
\hline\hline
\end{tabular}
\end{table}

\section{Discussion and Conclusions}
We have presented a collection of 250 new times of minima for 86 eclipsing binary systems, obtained between 2014 and 2024 using DSLR photometry. All timings were derived from uniformly reduced observations and determined using a consistent fitting and uncertainty estimation procedure, ensuring homogeneity across the entire dataset.
The results confirm that DSLR cameras can provide eclipse timings with sufficient precision for many practical applications in eclipsing binary research. While such detectors cannot fully match the performance of modern dedicated CCD and CMOS cameras, they remain a viable and accessible option for long-term photometric monitoring. We have also shown that useful eclipse timings can be obtained not only from dark-sky sites but also from locations affected by moderate light pollution, thereby broadening the potential contribution of amateur and semi-professional observers to this type of systematic monitoring.

The timings presented here extend the available observational coverage of the studied systems and may serve as useful input for future investigations of orbital period variations and other long-term phenomena in eclipsing binaries.

\acknowledgements
The Slovak Research and Development Agency supported this work under contract no APVV-24-0160.

\bibliography{demo_caosp310}

\end{document}